\documentclass{article}

\usepackage{arxiv}

\usepackage[utf8]{inputenc} 
\usepackage[T1]{fontenc}    
\usepackage{hyperref}       
\usepackage{url}            
\usepackage{booktabs}       
\usepackage{amsfonts}       
\usepackage{nicefrac}       
\usepackage{microtype}      
\usepackage{lipsum}
\usepackage{graphicx}

\usepackage{mathtools}
\usepackage{appendix}
\usepackage{amsmath}
\usepackage{amsthm}
\usepackage{amsfonts}
\usepackage{amssymb}
\usepackage{graphicx}
\usepackage[colorinlistoftodos]{todonotes}
\usepackage{color} 
\usepackage{subfigure}
\usepackage{ulem}
\usepackage{cancel}
\usepackage{soul}
\usepackage{tablefootnote}
\usepackage{comment}
\graphicspath{ {Images/} }

\newcommand{\nch}{n_{\textrm{ch}}}
\newcommand{\nco}{n_{\textrm{co}}}
\newcommand{\thch}{\theta_{\textrm{ch}}}
\newcommand{\thco}{\theta_{\textrm{co}}}
\newcommand{\alphach}{\alpha_{\textrm{ch}}}
\newcommand{\alphaco}{\alpha_{\textrm{co}}}
\newcommand{\Dth}{\Delta \theta}
\newcommand{\pch}{p_{\textrm{ch}}}
\newcommand{\pco}{p_{\textrm{co}}}
\newcommand{\pmean}{p_{\textrm{mean}}}

\graphicspath{ {./images/} }

\title{Modeling the evolution of collective synchrony}

\author{
 Guy Amichay* \\
  Department of Engineering Sciences and Applied Mathematics\\
  Northwestern Institute on Complex Systems\\
  National Institute for Theory and Mathematics in Biology\\
  Northwestern University\\
  Evanston, IL 60208 \\
  \texttt{guy.amichay@northwestern.edu} \\
   \And
 Ruoming Gong* \\
  Department of Engineering Sciences and Applied Mathematics\\
  Northwestern University\\
  Evanston, IL 60208 \\
  \texttt{ruoming.gong@northwestern.edu} \\
  \And
 Daniel M.~Abrams \\
  Department of Engineering Sciences and Applied Mathematics\\
  Northwestern Institute on Complex Systems\\
  National Institute for Theory and Mathematics in Biology\\
  Physics and Astronomy\\
  Northwestern University\\
  Evanston, IL 60208 \\
  \texttt{dmabrams@northwestern.edu} \\
}

\begin{document}
\maketitle
\begin{abstract}
Group synchrony in the animal kingdom is usually associated with mating. Being in sync is likely advantageous, as it may help in luring the opposite sex. Yet there are also disadvantages---such as the homogenization of the group---which make it harder for individuals to stand-out.  Here we address this tradeoff, bringing together the Kuramoto model with concepts from evolutionary game theory. We focus on the existence of self-interested ``cheaters,'' which have been extensively studied in a variety of species. In our scenario, cheating individuals take part in the synchronous group display but position themselves (in terms of phase) slightly ahead or behind the pack. This allows them to enjoy both the group benefit of advertisement and the individual benefit of being unique.  But a group can only tolerate a limited number of such individuals while still achieving synchrony.  We therefore incorporate a from of ``policing'' into our model: if an individual strays too far from the group's synchronous phase, they reveal themselves as dishonest and are punished. Our model offers testable predictions regarding natural population compositions, and will hopefully spur further investigation into not only how, but also why, natural systems synchronize. 
\end{abstract}


\section{Introduction}\label{sec:intro}

Synchrony plays a crucial role in the behavior of many animals, often manifesting as coordinated acoustic or visual signals \cite{greenfield2021rhythm}. In many cases, synchrony is thought to emerge due to the evolutionary advantages it provides, as opposed to simply being epiphenomenal. A well-known example is synchronization in mating displays\footnote{Collective courtship displays---also known as ``lekking''---need not be synchronous\cite{kirkpatrick1991evolution, pomiankowski1995resolution}.}, where coordinated flashing or calling can amplify signals, making them more effective for attracting mates over longer distances (the ``beacon hypothesis''\cite{buck1978toward}).

However, synchrony also occurs in contexts where its benefits are less obvious, raising intriguing questions. For instance, certain species of male fireflies exhibit synchronized flashing, which is thought to help attract females from afar \cite{buck1978toward}. Yet this raises a paradox: if all individuals flash simultaneously, how can any one of them stand out to a potential mate? Similar behavior is observed in fiddler crabs, where males wave their claws in synchrony, especially in the presence of females. Strikingly, some species even exhibit this behavior in the absence of females, suggesting that the function of synchrony may go beyond reproductive signaling. These examples highlight the need to better understand the evolutionary origins and stability of collective synchronization.

Empirically, getting traction on these questions poses multiple problems. Precise measurement of these behaviors in natural ecological contexts is challenging \cite{khatib2025, prasertkul2025}; automatically detecting where and when mating events occurr adds another layer of complication. Without such data, though, it is difficult to gain insight into which males are sexually selected for. 

We thus focus our efforts here on theory, with the hope that predictions will later be tested against field data. Mathematically, the Kuramoto model has long served as a foundational framework for studying synchronization in large populations \cite{kuramoto1975international,strogatz2000kuramoto}. When one accounts for the individual cost of synchronization, it becomes natural to frame the problem using the language of mean-field games \cite{hofer2025synchronization, Cesaroni2024, Antonioni2017, Tembine2014}. In this approach, each individual seeks to minimize a cost function, and the system settles into a Nash equilibrium that balances synchronization benefits and individual costs. However, standard mean-field game approaches usually fall short with regard to long-term trade-offs, as they only provide the Nash-equilibrium for the current population, and do not say anything about how the population will evolve. As such, the resulting equilibrium does not necessarily reveal which synchronization strategies are preferred or stable over evolutionary time.

Evolutionary game theory provides another lens for studying this problem \cite{Tripp2022, Smith1988}, allowing us to examine which synchronization strategies will be favored and adopted at equilibrium. This approach emphasizes the role of payoff structures in shaping strategy. Yet, in many cases, the choice of the payoff matrix lacks a principled justification, limiting the explanatory power of the framework.

Here we combine the Kuramoto model with an evolutionary game-theoretic framework to investigate the emergence and stability of synchrony in animal groups. Inspired by dishonest behavior (``cheating'') found in quorum sensing bacteria \cite{sandoz2007social}, fiddler crabs \cite{backwell2000dishonest} or even firefly ``femmes fatales" \cite{lloyd1965aggressive}, we imagine a scenario where cheaters exist within synchronous groups. Cheating, in this context, would mean taking part in the display, but positioning oneself as an outlier with respect to phase: that is, signaling with some phase advance or delay relative to the group. This would allow the cheaters to, on the one hand, appear as cooperative (they are taking part in generating the collective signal), but to also stand out---potentially giving them an advantage relative to the rest. We ask if, and how, they could play a part in the synchronous dynamics.   

Finally, we also introduce the concept of policing (found, e.g., in bacteria\cite{wang2015quorum} and honeybees\cite{halling2001worker}). Without policing, there may be no incentive to cooperate, so this is a key ingredient to stabilizing synchronous dynamics on an evolutionary time scale. As cheaters can hamper collective benefits, if other cooperating individuals detect such activity, they may ``enforce'' the rules (we don't specify here how this may be achieved; it could also be manifested differently in different species). As such, when cheaters become too obvious (e.g., they call anti-phase to the rest of the group) they become susceptible to detection and punishment.


\begin{figure}[t!]
    \centering

    \subfigure{\includegraphics[width=0.96\columnwidth]{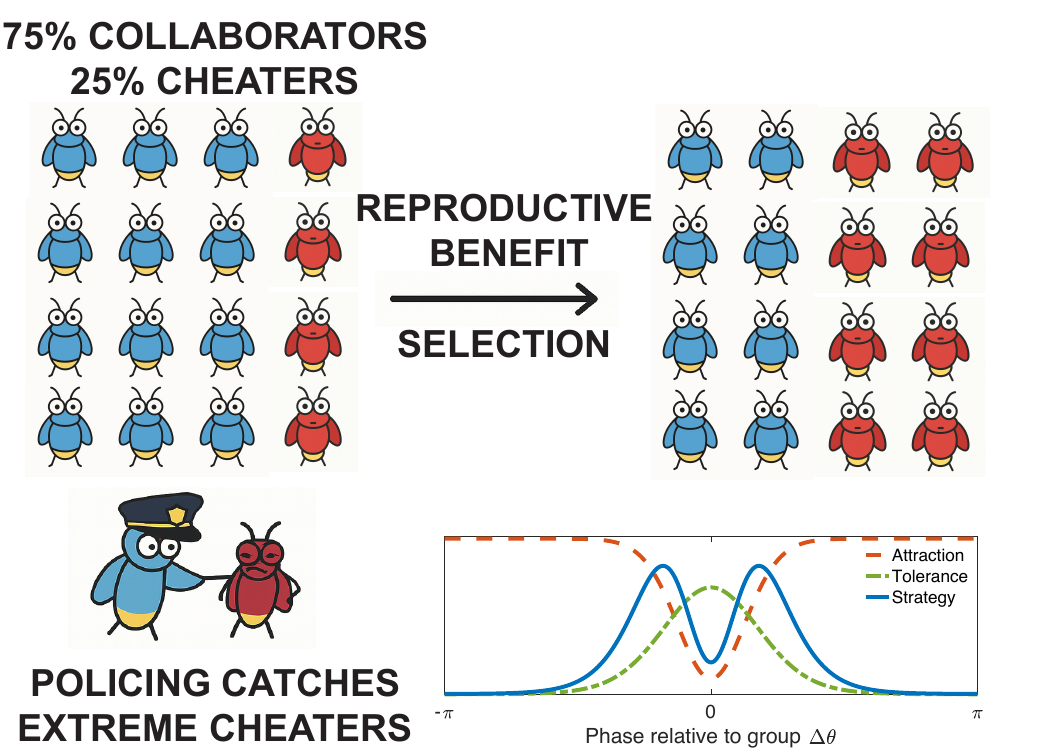}} 
    
    \caption{\textbf{Modeling approach.}
    Upper part shows a schematic that depicts our modeling process: in each generation the group size stays fixed, but the relative amount of cheaters may change due to selection. Note that we display fireflies in this example, but the model is relevant also to other similar systems such as fiddler crabs, frogs, crickets, etc.  The lower panel shows how individuals strategically act. The red dashed line is the ``attraction function,'' the green dot-dashed line is the policing ``tolerance function,'' and the blue solid line is the overall ``strategy function.'' In the mating context, individuals that deviate from the group have a higher chance to be selected, so the preference function reaches its minimum at $\Dth = 0$. By contrast, policing (the inverse of tolerance) targets individuals who deviate too far, leading the tolerance function to be maximized at $\Dth = 0$. The strategy function, proportional to the product of the preference and tolerance functions, illustrates the strategies that yield the highest individual benefit. }
    \label{fig:diagram}
\end{figure}

\section{Results}\label{sec:results}
\subsection{Models}

Our models incorporate two key time scales: a ``fast'' (within-generation) time scale representing the oscillatory behavior of an individual during their lifespan, and a ``slow'' time scale representing multi-generation evolutionary processes.

For the fast time scale, we assume that each individual's behavior follows the Sakaguchi-Kuramoto model\cite{sakaguchi1986soluble} (a modified Kuramoto model\cite{strogatz2000kuramoto}):
\begin{equation}
    \frac{d}{dt}\theta_i = \omega_i + \frac{K}{N}\sum_{j=1}^{N} \sin(\theta_j - \theta_i + \alpha_i), \label{eq: Kuramoto}
\end{equation}
where $\theta_i$ is the phase of individual $\omega_i$ is its intrinsic frequency (which we take to be universal), $K$ is the coupling strength, and $\alpha_i$ represents the individual's phase lag strategy. For simplicity, we assume that the strategy of each individual is fully characterized by its phase lag parameter $\alpha_i$.  A phase lag of zero would mean that the individual dynamics are geared towards in-phase relationships with neighbors. A nonzero phase lag pushes the individual to be ahead or behind its neighbors in terms of phase (though  system with identical individuals with nonzero phase lag, e.g., could still yield synchrony).  Phase lags were considered in prior works \cite{yeung1999time}, but were treated as a proxy for delays; here we reinterpret the meaning, and suggest that this acts as a built-in strategy to aim to be different. 

On the slow time scale, we model natural selection as optimizing a net ``payoff'' to each individual.  That payoff is dependent on both group-level and individual level properties, and is expressed as the difference between a benefit and a cost.  The benefit $b_i$ for individual $i$ is defined as the product of two terms: $g_i$, the expected number of females attracted to the group (determined by group-level synchrony), and $f_i$, an individual ``strategy function,'' which quantifies the value of an individual's flash timing strategy. We assume $g_i = g_i(R)$ is a positive increasing function of the order parameter $R$ defined as 
\begin{align}
    R e^{i\phi} = \frac{1}{N}\sum_{j=1}^{N} e^{i\theta_j},
    \label{eq:order_parameter}
\end{align}
where $R \in [0,1]$ quantifies the level of synchrony, and $\phi$ is the average phase. The strategy function $f_i = f_i(\theta_i - \phi)$ is a function of phase difference $\theta_i - \phi$ \footnote{This is the phase of oscillator $i$ relative to the mean group phase, and we will later denote it by $\Dth$.}, and is taken to be a product of an ``attraction'' and a police ``tolerance'' function, as illustrated in Fig~\ref{fig:diagram}. It captures both how individual timing relative to the group influences mate attraction, and how standing out from the group might be tolerated by policing\footnote{Policing enables the group to act against extreme (or obvious) cheaters, reducing their payoffs as a result.}:
\begin{align}
    b_i = g(R)f_i\left(\theta_i - \phi\right) =  g(R) \left[\textrm{att}(\theta_i - \phi) \textrm{tol}(\theta_i - \phi) \right].
    \label{eq: individaul benefit}
\end{align}

On the cost side, letting
\begin{align}
    u_i = \frac{K}{N}\sum_{j=1}^{n} \sin(\theta_j - \theta_i + \alpha_i)
\end{align}
denote the active control or effort\footnote{assuming that oscillating ($\omega$) isn't part of the effort, as they are all oscillating equally; so here we only address the effort to adjust relative to others.} exerted by individual $i$, we define the individual cost $c_i$ over a time window $[0,T]$ as
\begin{align*}
    c_i = \frac{1}{T}\int_{0}^{T} |u_i(t)|^2 dt.
\end{align*}

The net payoff $p_i$ is defined as the difference between the benefit and the cost, weighted by a relative cost parameter $\beta$:
\begin{align}
    p_i = b_i - \beta c_i.
    \label{eq:payoff}
\end{align}


\subsection{Group membership dynamics with binary phase lag}

We begin with the simple scenario consisting of two groups of individuals, cooperators and cheaters, each of which comprises a corresponding fraction of the population $\nco$ and $\nch$ (with $\nco + \nch = 1$).  Then the evolutionary dynamics corresponding to the above payoff function Eq.~\eqref{eq:payoff} satisfies:
\begin{equation}
\begin{aligned}
    \nco^{(g+1)} &= \nco^{(g)} \left[1 + k_1(\pco - \pmean)\right], \\
    \nch^{(g+1)} &= \nch^{(g)} \left[1 + k_1(\pch - \pmean)\right], 
\end{aligned}
\label{eq:discrete_evolution}
\end{equation}
where the index $g$ indicates the generation number and $k_1$ quantifies the strength of the linear dependence between individual payoff and reproduction chances. (See Appendix \ref{section: two_group equations} for the derivation of Eq.~\eqref{eq:discrete_evolution}) \
 
The population evolution in Eq.~\eqref{eq:discrete_evolution} represents expected values.  It can be implemented with stochastic effects by defining the relative reproductive rate $r_i$ to follow the Fermi selection rule \cite{PERC2010109} as a function of payoff, 
\begin{align}
    r_i \propto \frac{e^{p_i}}{\max_{j}e^{p_j}}, \label{eq: reproduction}
\end{align}
so that each individual in the next generation has a probability of $r_i \big/ \sum_j r_j$ to be the offspring of individual $i$.


Taking $k_1 = k \Delta t$, where $k$ now captures the time scale, we set $n^{(g)} \mapsto n(t)$, and $n^{(g+1)} \mapsto n(t+\Delta t)$ and let $\Delta t \to 0$ to obtain the continuous version of Eq.~\eqref{eq:discrete_evolution}:
\begin{equation}
\begin{aligned}
    \frac{d}{dt}\nco &= k \nco(\pco - \pmean), \\
    \frac{d}{dt}\nch &= k \nch(\pch - \pmean),
\end{aligned}
\label{eq:cts time model}
\end{equation}
where $\pmean = \nco \pco + \nch \pch$.


We numerically explore Eqs.~\eqref{eq:cts time model} with the interpretation of cooperators as individuals with a phase lag of $\alphaco = 0$ and cheaters as individuals with a phase lag $\alphach \neq 0$ (but equal for all of them---they are identical). We emphasize that here $\alphaco$ and $\alphach$ are fixed: selection is applied only to the abundances of group memberships. Another way to view this is that the $\alpha$ values \emph{can change}, but these changes are restricted to exactly two specific values (namely, zero and $\alphach$).  

\begin{figure}[p]
    \centering
    \subfigure{\includegraphics[width=0.7\columnwidth]{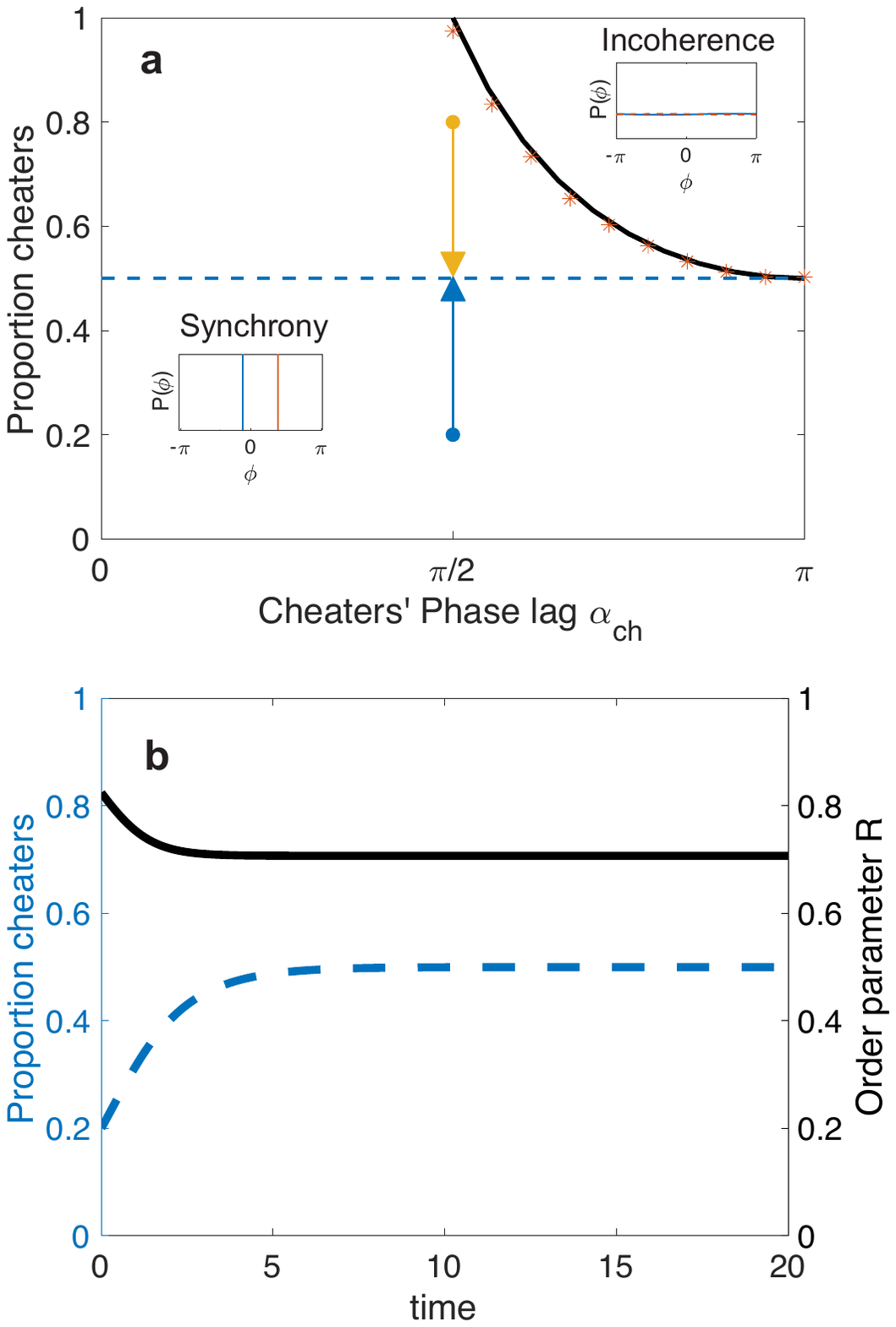}}
    \caption{\textbf{Phase transition for two-group model.} Panel (a) illustrates a simplified case where the population consists of two distinct types of individuals: cooperators ($\alphaco=0$) and cheaters, for fixed cheater phase lags $\alphach$ from $0$ to $\pi$. Under uniform coupling strength and natural frequency, the system exhibits an incoherent state when $\alphach \geq \pi/2$. The theoretical transition boundary (solid black curve) is given by the expression $(1-\cos(\alphach))^{-1}$ while red dots represent results from numerical simulations. The blue/yellow dots in panel (a) represent the initial population sizes of cheaters, while the dashed blue line shows the equilibrium population of cheaters as a function of $\alphach$. Insets in panel (a) show states commonly observed in the long-term behavior at synchrony and incoherence. Panel (b) demonstrates the population dynamics driven by competition between cooperators and cheaters---specifically the dynamics for blue arrow in panel (a). Here cheaters (blue dashed line) adopt a phase lag of $\alphach = \pi/2$. The initial condition is set to $\nch=0.2$, representing the initial cheater fraction. }
    \label{fig:phase transition} 
\end{figure}

Fig.~\ref{fig:phase transition}(a) presents the results of numerical simulations with $N=1000$ oscillators and illustrates the critical threshold beyond which a population of cheaters can no longer sustain synchrony.  Fig.~\ref{fig:phase transition}(b) shows the dynamical relaxation to equilibrium in a typical case. For persistent populations, the proportion of cooperators and cheaters at equilibrium is always $50-50$ (as showed in Fig.~\ref{fig:phase transition}(a)). This is due to the fact that the minority population loses its advantage as it grows to become half the population.

Simple analytical results can be obtained in this case. Assuming that at equilibrium each subpopulation is synchronized and all oscillators are frequency locked: cooperators share the same phase $\thco$ and cheaters share the same phase $\thch$, with $\Dth_0 = \thch-\thco$ constant over time. $\nco$ and $\nch$ are the fractions of cooperators and cheaters in the population respectively. Substituting these into Eq.~(\ref{eq: Kuramoto}), we obtain:
\begin{equation}
\begin{aligned}
    \frac{d}{dt}\thco &= \omega + k \nch \sin(\Dth_0) \\
    \frac{d}{dt}\thch &= \omega + k \left[ \nco \sin(-\Dth_0 + \alphach]) + \nch \sin(\alphach) \right]\;,
\end{aligned}
\label{eq: binary frequency}
\end{equation}
or, expressed fully in the angle difference variable $\Dth_0$,
\begin{equation}
    k^{-1} \frac{d}{dt}\Dth_0 =  \nco\sin(-\Dth_0+\alphach) + \nch\sin(\alphach) - \nch \sin(\Dth_0).    
\end{equation}
Setting the time derivative to zero, we find the equilibrium solution 
\begin{align*}
    \Dth_0^* = \alphach\;.
\end{align*}
Linear stability analysis reveals that this solution is stable when
\begin{equation}
    \nch < \frac{1}{1-\cos(\alphach)},   
\end{equation}
which matches the numerical results shown in Fig.~\ref{fig:phase transition}(a). In Appendix \ref{section:Linear stability for binary} we generalize these results to any binary choice of $\alpha$.

The key message we take away from this model: It is possible for both cheaters and cooperators to coexist in the population, but there is an upper limit on the number of cheaters the population can sustain. That upper limit is $1 / (1 - \cos(\alpha_{ch}))$.


\subsection{Evolutionary dynamics with phase lag distributions}

In this section, we introduce a new aspect: \textit{mutation}. The inclusion of mutations is to make the evolutionary dynamics more realistic now that we relax the binarization of the phase lags---now $\alpha$ may change across generations. Due to this modification, we need to include the notion of policing. Without policing, everyone in the group will eventually cheat to increase their reproductive success. As a result, the group won't be able to synchronize, which in turn will lead the population to die out.

\begin{figure*}[t]
\centerline{\includegraphics[width=\textwidth]{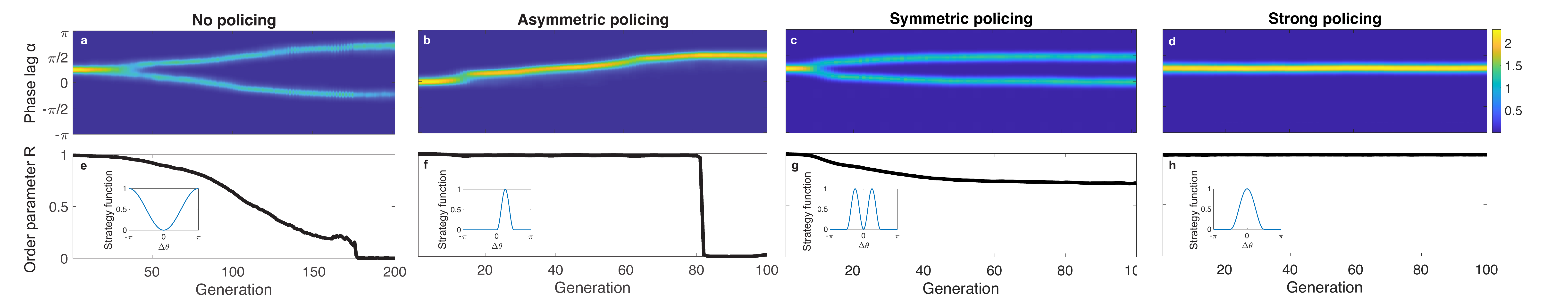}}
\caption{\textbf{Evolutionary dynamics with continous phase lag distribution.} Heatmaps in upper panels show probability density for each phase lag $\alpha$ over generations; lower panels illustrate evolution of the corresponding order parameter $R$. Panels (a) and (e): no effective policing and initial condition (IC) is $\mathcal{N}(\pi/4, 0.1)$. Panels (b) and (f): asymmetric policing, IC is $\mathcal{N}(0, 0.1)$. Panel (c) and (g): symmetric policing function, IC is  $\mathcal{N}(\pi/4, 0.1)$.   Panels (d) and (h): strong policing function, IC is  $\mathcal{N}(\pi/4, 0.1)$. Insets in lower panels show the overall strategy function resulting from a fixed symmetric attraction function and the given policing (see Fig.~\ref{fig:diagram}(b)).  All trials had $N$=1000 oscillators. 
}
\label{fig:population_dynamcis_pi4}
\end{figure*}

With policing, we allow the strategy function $f(\Dth)$ to incorporate information beyond mere attractiveness: it also now indicates the optimal balance between maximizing mate attraction and minimizing police intervention (due to, e.g., female choice constraints or actions by competitive males). We note that the payoff primarily depends\footnote{Since payoff $p_i = b_i - \beta c_i = f_i(\Dth) g(R) - \beta c_i$ where $g(R)$ is uniform over the population and $c_i$ is nearly uniform over the population.} on the choice of the strategy function $f(\Dth)$, which itself depends only on deviation from the mean phase. Other components of the payoff are approximately uniform across individuals at equilibrium due to frequency locking.




We explore the dynamics of the evolution of the population's $\alpha$ distribution numerically. For each generation, to compute individual payoffs at equilibrium, we numerically integrate the dynamics given by Eq.~\eqref{eq: Kuramoto}. After transients have decayed, we use the final phases to compute the payoff of each individual. The reproduction probability for each individual is determined by Eq.~\eqref{eq: reproduction} given the payoff. 

To include mutation, a random subset of the new population undergoes mutation on the phase lag $\alpha$ such that:
\begin{align}
    \alpha_i^{(\text{offspring})} = \alpha_i^{(\text{parent})} + I_i \epsilon_i,
    \label{eq:mutation}
\end{align}
where
\begin{align*}
    I_i = 
\begin{cases}
1 & \text{if } i \in \text{mutation set} \\
0 & \text{otherwise}, 
\end{cases}
\end{align*}
and $\epsilon_i \sim \mathcal{N}(0, \sigma_{\text{mutate}})$ represents the mutation perturbation.

In Fig.~\ref{fig:population_dynamcis_pi4}, we initialize the population's phase lag distribution as a normal distribution with mean $\mu$ and standard deviation $\sigma$. We consider four cases, all under the assumption that the attracted individuals have no preference for phases advanced versus delayed relative to the group (symmetric attraction function). The four scenarios differ in how policing is applied through the tolerance function: (1) no policing (without any tolerance function), (2) asymmetric policing (e.g., only individuals lagging behind the group are penalized), (3) symmetric policing (individuals both ahead of and behind the group are penalized), and (4) strong policing (individuals even slightly away from the mean group phase are penalized to such an extent that there is no net benefit to being out of sync, despite mate attraction being minimized when in sync). 

Fig.~\ref{fig:population_dynamcis_pi4}(a) shows results for the ``no policing'' scenario. There individuals are incentivized to cheat as much as possible to maximize their individual benefit. This leads to a breakdown of coordination, and the population ultimately goes extinct due to the resulting low level of synchrony. 

Fig.~\ref{fig:population_dynamcis_pi4}(b) shows results for asymmetric policing, and more specifically, for the case where only individuals lagging behind the group are penalized. Here the distribution of phase lags $\alpha$ drifts in the positive direction, with the mean eventually approaching $\pi/2$, when a breakdown of synchrony occurs and the population goes extinct.

In Fig.~\ref{fig:population_dynamcis_pi4}(c), symmetric policing is imposed on both individuals ahead and behind the group.  We observe that the population splits into two clusters and stabilizes. 

Interestingly, this symmetry in the policing function appears to be necessary for survival of the population.  Asymmetry leads eventually to takeover by cheaters, and thus populations that survive, according to this model, must find a way to impose a symmetric cost on either leading or lagging cheaters.

Finally, Fig.~\ref{fig:population_dynamcis_pi4}(d) illustrates the case of strong policing 
, where the incentive to cheat is entirely eliminated: the cost of deviating from the group outweighs any potential benefit. In this regime, unsurprisingly, the population remains stable and synchronous; the system is very similar to the case with no cheaters.

We note that the long-term survival of the population depends on the interplay between the choice of the policing function and the initial conditions $\mu, \sigma$. If the policing is too weak (i.e., the gap between the peaks in the strategy function is too wide) and/or the initial population is not collaborative enough (i.e., $\mu$ is too far from 0), the population will fail to survive---see Appendix \ref{section:IC condition} for exact limits.

Additional comments on the robustness of our results can be found in Appendix \ref{sec:robustness}.

\section{Discussion and Conclusions}


We have developed a novel approach based on the Sakaguchi-Kuramoto model where we swap the usual heterogeneity of natural frequencies for a different source of irregularity: a distribution of phase lags $\alpha$ (usually taken to be a constant, although see some exceptions: \cite{smirnov2024dynamics, pikovsky2024dynamics, pazo2011kuramoto, vlasov2014synchronization, lee2009large, izhikevich1998phase}). Phase lags can take on different interpretations, including individual delays (e.g., due to computation times internal to each agent) or, as we've considered here, purposeful attempts at misalignment with others. 
Since multiple species are known to synchronize while lekking (e.g., fireflies\cite{buck1966biology}, crickets\cite{greenfield1993katydid}, frogs \cite{aihara2009modeling} and crabs \cite{backwell2019synchronous}---see \cite{greenfield2021rhythm} for a review on the topic), we think our model has the potential for wide application.

Our main result is the discovery that populations of both cooperators and cheaters can coexist even in a highly simplified model, and we analytically derive an upper limit on the fraction of cheaters. This suggests that natural animal collectives might indeed contain a subset of cheaters. In addition, according to our model, symmetric policing---where straying ahead or behind the pack is equally enforced---is the only stable option with the presence of cheaters. Thus if policing is somehow observed in a natural system, we make the prediction that it will be equally applied to both early and late signalers. 




The structural instability inherent in this symmetric system (i.e., slight perturbations to a symmetric policing function would lead to extinction) can potentially be resolved by introducing additional stabilization mechanisms. For example, over evolutionary time scales, only populations with symmetric payoff structures may persist, while those lacking this symmetry are driven to extinction.

It is difficult to distinguish cheaters from cooperators based solely on the phase distribution, as cheaters tend to form one synchronized cluster while cooperators form another in the steady state. This may be because both groups incur the same coupling cost in our model, given the universal coupling strength $K$. However, if cheaters and cooperators experience different coupling costs, it may become possible to differentiate them from the phase distribution, as suggested in \cite{Hong2011}.


One limitation of our study is that we use a global order parameter $R$ to quantify synchrony, which does not account for clusters of phases. For example, when the population splits into two phase clusters, a perfect cooperator may not respond to the global mean phase, but rather to the mean phase of their local cluster. 

In addition to cheating and policing, other mechanisms may contribute to the collapse of synchrony in some species while preserving it in others. Selective attention and local avoidance strategies have been observed in frog choruses, which can also disrupt full synchronization \cite{aihara2019frogmodel, ota2020frogchorus}.

Future work could focus on generalizations of our model. For simplicity we kept the model in an all-to-all setup. Of course the positions of the males in space (whether in 1-, 2- or 3D) could be important. It is easy to imagine that the relative positioning of cheaters, be it within clusters or more uniformly distributed across the group, could influence the overall dynamics.

We conclude with a hope that our model could offer testable predictions for real animal populations. To that end, carefully defining certain individuals as cheaters is key. Ideally, when analyzing pairwise phase differences, some nearby neighbors will appear to have consistent nonzero phase differences\footnote{Note that nonzero phase differences could also arise simply due to distance---minimal imperfections between neighbors can accumulate causing a rippling affect such that farther neighbors may appear to have some larger phase differences.}. If indeed cheaters are well defined and detected, then a rigorous quantification of their fraction within the group could be assessed, yielding insight into the real-world utility of our model.   


\section{Methods}
\subsection{Binary phase lag simulation}
We start with the simple case where we only allow the phase lag to take on two given values: $\alphaco$ and $\alphach$, for cooperators and cheaters respectively. The population sizes of cooperators and cheaters evolve according to Eq.~\eqref{eq:cts time model}. The simulation results are deterministic given the initial conditions.   

We use the following functions to calculate the payoff:
\begin{align*}
    &g(R) = R,\\
    &f(\Dth) = 1 - \cos(\Dth),
\end{align*}
where $\Dth = \theta - \phi$. These are simple choices consistent with our assumptions that $g(R)$ be a monotonically increasing function of $R$ and $f(\Dth)$ be a $2 \pi$ periodic function that penalizes individuals with small phase deviations. 

There are two possible equilibrium states of the population. When $\nch < (1-\cos\alphach)^{-1}$, all cooperators have the same phase $\theta_{co}$ and all cheaters have same phase $\theta_{ch}$ with $\theta_{ch} - \theta_{co} = \alphach$ at equilibrium. In this case, $\pco$, $\pch$, and $\pmean$ can be calculated explicitly as
\begin{align*}
    \pco &= R (1 - \cos(\phi)),\\
    \pch &= R (1 - \cos(\alphach - \phi)),\\
    \pmean &= \nco \pco + \nch \pch,
\end{align*}
with $\exp(i\phi) = \nch \exp(i\alphach) + \nco$ ($\phi$ defined as in Eq.~(\ref{eq:order_parameter})). 

Note that we are free to remove the cost term from the payoff here, since the cost is the same for all individuals in this scenario, and thus removing it does not change the value of $(\pco - \pmean)$ or $(\pch - \pmean)$.

We do not simulate the case when $\nch > (1-\cos\alphach)^{-1}$ since the order parameter $R$ approaches zero in this regime, indicating that the population cannot maintain synchrony and thus lacks sufficient fitness for reproduction, ultimately leading to extinction.

To numerically determine the critical boundary, we integrate Eq.~\eqref{eq: Kuramoto} using various initial population compositions and values of $\alphach$. We sweep $\alphach$ from $0$ to $\pi$. For each value of $\alphach$, we conduct a series of simulations with increasing initial fractions of cheaters $\nch(t=0)$. The critical proportion of cheaters $\nch^\textrm{(crit)}$ is identified as the smallest cheater fraction that causes the system to become incoherent at equilibrium. This procedure yields a series of $(\alphach, \nch^\textrm{(crit)})$ pairs and, as shown in Fig.~\ref{fig:phase transition}, they appear to follow the curve $\nch^\textrm{(crit)} = (1-\cos\alphach)^{-1}$, consistent with the analytical prediction. 

\subsection{Population dynamics simulation with mutation}
In these simulations we allow the individual phase lags to mutate across generations. A key practical distinction is that we can obtain a closed-form expressions for equilibrium payoff in the binary case, but we cannot do so in the model with mutation: payoff needs to be evaluated numerically. For each generation, we numerically integrate the dynamics given by Eq.~\eqref{eq: Kuramoto} up to $T_\textrm{final} = 100$ with $N = 1000$ individuals. Without loss of generality, we choose $\omega = 0$ and $K = 1$.

For payoffs with symmetric policing, we choose 
\begin{align*}
    & g(R) = R, \\
    & f(\Dth) = 
    \begin{cases} \cos(4(\Dth - \frac{\pi}{4})) + 1, & |\Dth| \mathrel{<} \frac{\pi}{2} \\
    0, & |\Dth| \geq \frac{\pi}{2}
    \end{cases}.
\end{align*}
For payoffs with asymmetric (one-side) policing, we choose 
\begin{align*}
    & g(R) = R, \\
    & f(\Dth) = 
    \begin{cases} \cos(4(\Dth - \frac{\pi}{4})) + 1, & |\Dth| \mathrel{<} \frac{\pi}{2} \\
    0, & |\Dth| \geq \frac{\pi}{2}
    \end{cases}.
\end{align*}
In both cases we set $\beta = 1$ to calculate the payoff.  The choice of $\beta$ is not critical as cost $c_i$ is nearly identical across individuals. The payoff for each individual is calculated with Eq.~\eqref{eq:payoff} averaged over the final $10\%$ of the simulation time.

In our simulations, we randomly select $10\%$ of the population to mutate at each generation and model those mutations via additive perturbations of magnitude $\epsilon_i \sim \mathcal{N}(0, \sigma)$ as shown in Eq.~\eqref{eq:mutation}.  We use $\sigma = 0.05$, since smaller $\sigma$ helps keep the resultant dynamics smooth. The magnitude of these perturbations controls the rate at which the distribution of $\alpha$ drifts across generations.

\bibliographystyle{plain}
\bibliography{evolutionrefs} 


\newpage
\appendix
\appendixpage

\section{Derivation of two-group equations}\label{section: two_group equations}
\vspace*{12pt}
We assume that the expected number of offspring for an individual with payoff $p_i$, in a population with mean payoff $\pmean$, is linearly proportional to that difference $p_i - \pmean$.  So after $g$ generations, if there are $N_i^{(g)}$ individuals descended from an initial progenitor $i$, the number of new offspring (in expected value) will be
\begin{equation}
    N_i^{(g+1)} = (1-\gamma) N_i^{(g)} + k_0 N_i^{(g)}(p_i - \pmean),
    \label{eq: discrete population evolution}
\end{equation}
where $k_0$ is a constant of proportionality for birth and $\gamma$ determines the fraction of individuals who die in a single generation timescale.

Rewriting in terms of the fractional population $n_i^{(g)} = N_i^{(g)}/N^{(g)}$ where $N^{(g)} = \sum_i N_i^{(g)}$, we get
\begin{equation}
    n_i^{(g+1)} N^{(g+1)} = n_i^{(g)} N^{(g)} \left[1 - \gamma + k_0(p_i - \pmean)\right].
    \label{eq: discrete population evolution2}
\end{equation}
To enforce that $\sum_i n_i^{(g)} = 1$ in all generations (so $n$ represents a fraction as intended), the ratio $N^{(g+1)} /  N^{(g)} = 1-\gamma$ is required. Then this simplifies to
\begin{equation}
     n_i^{(g+1)} = n_i^{(g)} \left[1 + \frac{k_0}{1-\gamma}(p_i - \pmean)\right].
    \label{eq: discrete population evolution3}
\end{equation}
We can define $k_1 = \frac{k_0}{1-\gamma}$ as an effective generational growth rate for each subpopulation $i$.

In the simple case where there are only two types of individuals present in the population, this simplifies to two coupled equations.  Considering the two types as cooperators and cheaters, the system becomes
\begin{equation}
\begin{aligned}
    \nco^{(g+1)} &= \nco^{(g)} \left[1 + k_1(\pco - \pmean)\right], \\
    \nch^{(g+1)} &= \nch^{(g)} \left[1 + k_1(\pch - \pmean)\right], 
\end{aligned}
\end{equation}
where $\nco, \nch$ represents the fraction of the population of cooperators and cheaters respectively, and the index $g$ indicates the generation number.

\section{Linear stability analysis of the binary phase lag model}\label{section:Linear stability for binary}
\vspace*{12pt}
Assume at equilibrium, the whole population is frequency locked. The cooperators all have phase $\thco$, and the cheaters all have phase $\thch$, with $\Dth_0 = \thch-\thco$ constant over time. $\nco, \nch$ is the percentage of cooperators and cheaters in the population respectively ($\nco+\nch = 1$). Without loss of generality, we can set $\omega = 0$ and $K = 1$. Then we have (in a co-rotating frame)
\begin{align*}
    & \dot{\theta}_{co} = \nch \sin(\thch - \thco + \alphaco) - \nch\sin(\alphach),\\
    & \dot{\theta}_{ch} = \nco \sin(\thco - \thch + \alphach) - \nco \sin(\alphaco).
\end{align*}
The two eigenvalues of the Jacobian matrix for this system are:
\begin{align*}
    \lambda_1 = 0, \quad \lambda_2 = -\nco\cos(\alphaco) - \nch\cos(\alphach).
\end{align*}

Note that $\lambda_1$ is the eigenvalue corresponding to the direction $\theta_{ch} - \theta_{co} = \alphach - \alphaco$, i.e., perturbations that preserve the phase difference between cheaters and cooperators; the system is neutral to any perturbation in this direction, which simply reflects the freedom of choice of phase coordinate origin.

Setting $\lambda_2 < 0$, we find
\begin{align}
    \nch < \frac{\cos(\alphaco)}{\cos(\alphaco)-\cos(\alphach)}.
    \label{eq: two cluster general}
\end{align}

In particular, when $\alphaco = 0$ and $\alphach = \alpha$, this reduces to the stability condition
\begin{align*}
    \nch < \frac{1}{1 - \cos(\alpha)}.
\end{align*}
This sets an upper bound on the proportion of cheaters that may be present in a synchronous population.

\subsection{Connection between policing and initial cooperativeness}\label{section:IC condition}
Here we comment on the connection between the strategy function and the initial population distribution for long-term stability of the population.

Suppose $\sigma_{\text{mutate}}$ is small and the gap between the two modes of the strategy function is $\ell$. We observe that the population eventually stabilizes into two distinct subpopulations. The separation between the two modes of the strategy function approximately matches the phase lag difference between the two stabilized subpopulations, as individuals maximizing their benefits. Because the mutation process has zero mean (no directional bias for $\alpha$), the two subpopulations are expected to be equal in size and symmetrically located about the initial $\alpha$ distribution's mean $\mu$ (without lose of generality we choose $\mu>0$).
That is, $\mu_1 = \mu - \ell/2$ and $\mu_2 = \mu + \ell/2$. We focus on the case where $\mu_1 \in (-\pi/2, \pi/2), \mu_2 \in [\pi/2, 3\pi/2]$, which corresponds to parameter regimes in which the population can persist in the long term.

Assuming that the standard deviations of the two subpopulations are small, the stability condition can be approximated using Eq.~\ref{eq: two cluster general}, with $\alphaco = \mu_1, \alphach = \mu_2, \nch = 1/2$. This leads directly to the condition:
\begin{equation}
    \frac{\cos(\mu - \frac{\ell}{2})}{\cos(\mu - \frac{\ell}{2}) - \cos(\mu + \frac{\ell}{2})} \gtrsim \frac{1}{2}, \label{eq: IC condition}
\end{equation}

\section{Robustness}\label{sec:robustness}

\subsection{Robustness: discrete vs.~continous evolution}
\begin{figure}[ht!]
    \centering
    \includegraphics[clip, trim=10mm 3mm 15mm 10mm, width=0.6\columnwidth]{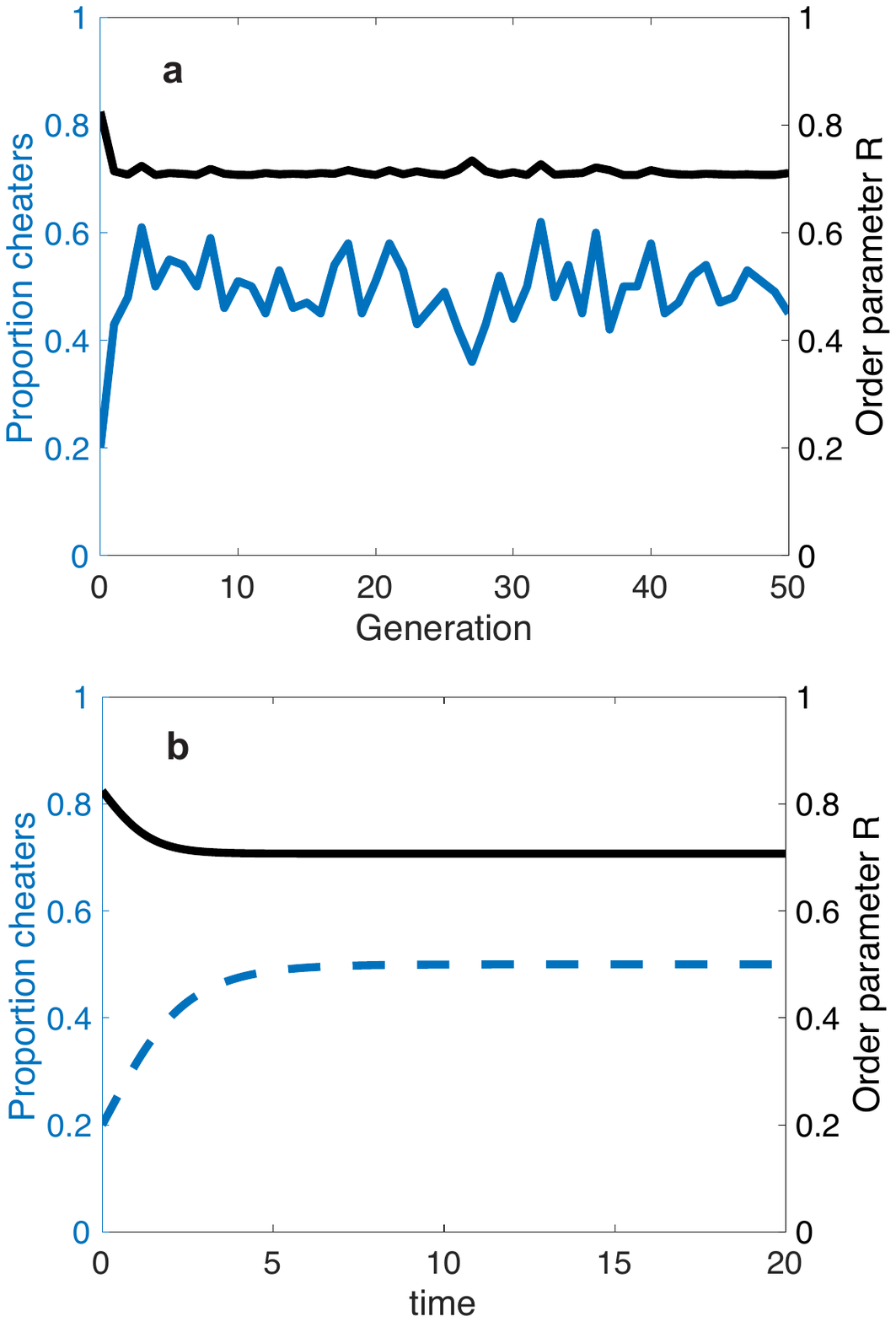}
    \caption{\textbf{Comparing discrete and continuous evolution equations.} Cooperators are characterized by a phase lag of $\alphaco = 0$, while cheaters adopt a phase lag of $\alphach = \pi/2$. When cooperators are the majority, cheaters gain a reproductive advantage by deviating from the mean phase of population. Conversely, when cheaters become dominant, cooperators---now the minority---regain a fitness advantage. This interplay creates cyclic fluctuations in population composition, oscillating around a mean of $50\%$. When population changes occur smoothly, the dynamics stabilize at an even split between cooperators and cheaters. }
    \label{fig:cooperator_cheater_competition}
\end{figure}

For the model with binary $\alpha$, we check that results from simulations with discrete generations (Eq.~(\ref{eq:discrete_evolution})) are consistent with those from equivalent simulations with continuous evolution (Eq.~(\ref{eq:cts time model})). In Figure \ref{fig:cooperator_cheater_competition}, as expected, the discrete simulation appears as a noisier version of the continuous one, with the overall picture remaining consistent.

\subsection{Robustness: asymmetric bimodal strategy function}
\begin{figure}[ht!]
    \centering
    \includegraphics[clip, trim=10mm 3mm 15mm 7mm, width=0.7\columnwidth]{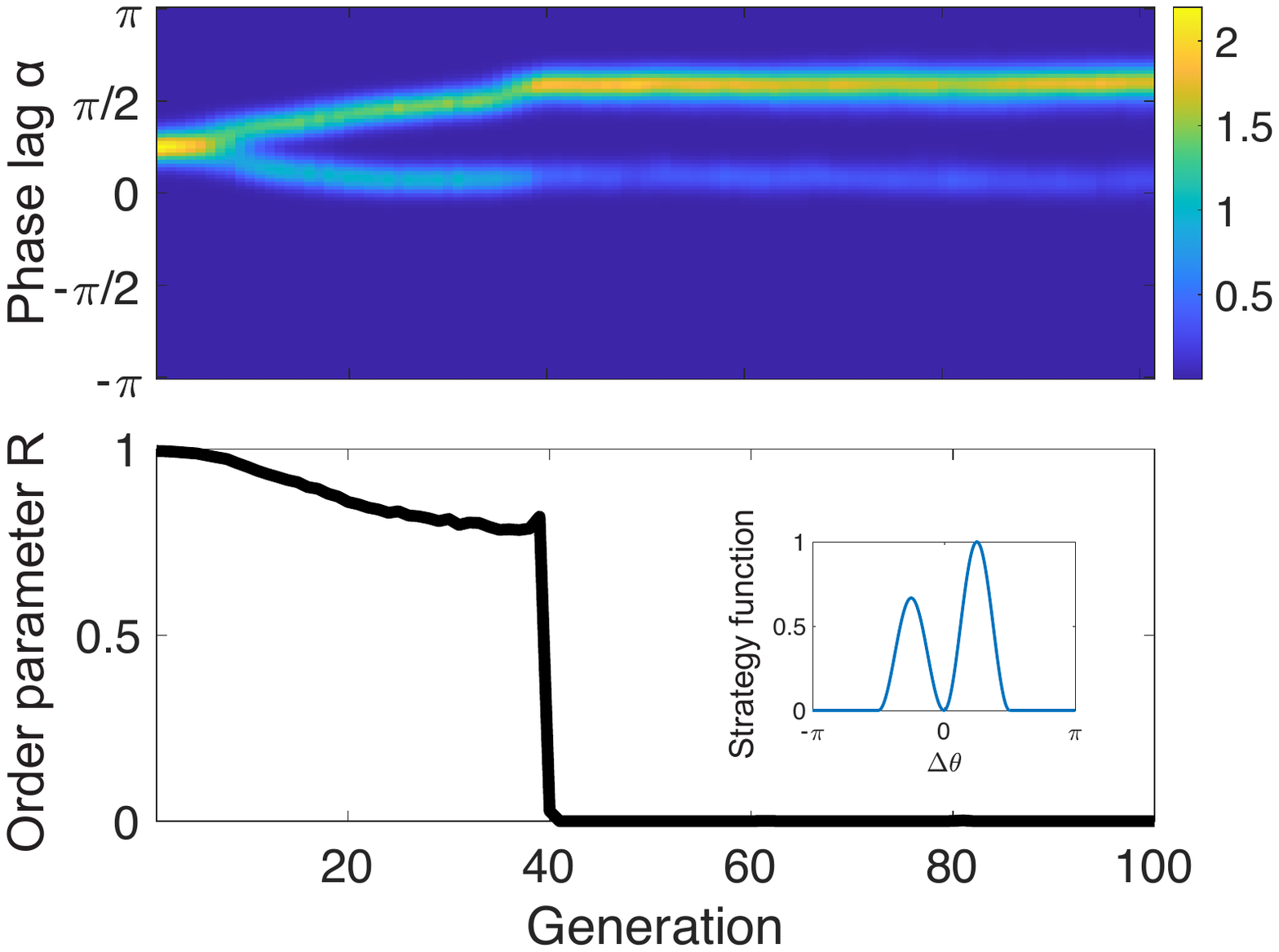}
    \caption{\textbf{Asymmetric policing.} When policing is asymmetric, it yields an asymmetric strategy function that rewards individuals ahead of the pack differently from those behind. In such cases, the population cannot sustain itself over the long term and eventually goes extinct. }
    \label{fig:population_dynamcis_asymmetry} 
\end{figure}

We examine numerically whether a symmetric choice of the policing function is essential for the long-term survival of the population. Specifically, in Fig.~\ref{fig:population_dynamcis_asymmetry}, we consider an asymmetric bimodal strategy function and observe that the subpopulation with negative phase lag eventually dies out due to the accumulated advantage of those with positive phase lag. As a result, the entire population fails to achieve synchronization.

\subsection{Robustness: nonzero payoff for perfect cooperators}
\begin{figure}[ht!]
    \centering
    \includegraphics[clip, trim=10mm 3mm 15mm 7mm, width=0.7\columnwidth]{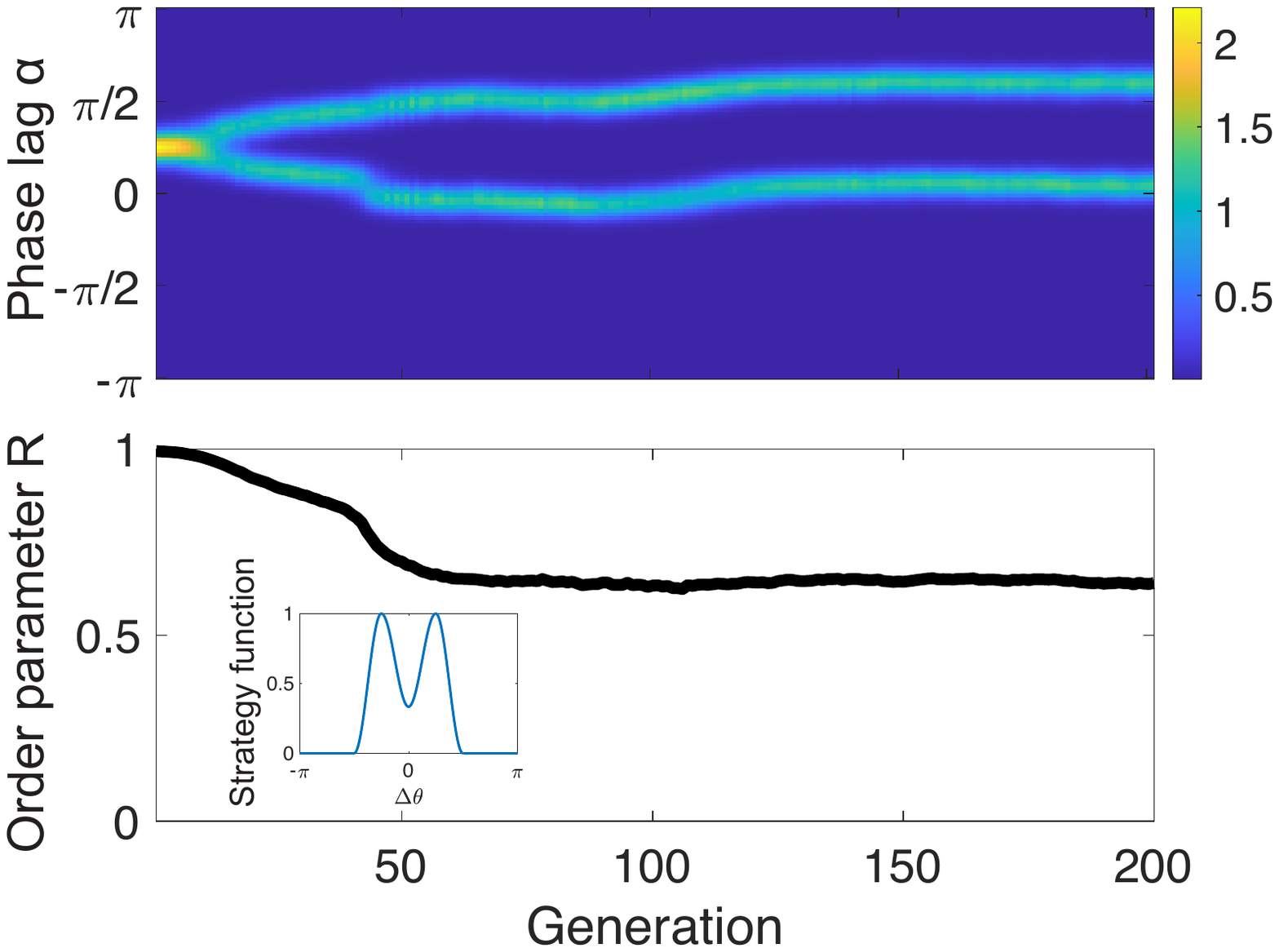}
    \caption{\textbf{Nonzero payoff at $\alpha=0$.} To test the robustness of our choice to have strategy function drop to zero at $\Dth=0$, we simulate an alternative strategy function with a non-zero value at $\Dth = 0$, as shown in the inset. The resulting population dynamics remained qualitatively unchanged as compared to Fig.~\ref{fig:population_dynamcis_pi4}(c), suggesting that the model is not sensitive to the specific value of the strategy function at $\Dth = 0$, as long as the overall symmetry is preserved.}
    \label{fig:population_dynamcis_nonzeroorg} 
\end{figure}

We examine the robustness of the simulation results when perfect cooperators ($\alpha = 0$) receive a nonzero payoff. As shown in Fig.~\ref{fig:population_dynamcis_nonzeroorg}, the results exhibit the same qualitative behavior as those in Fig.~\ref{fig:population_dynamcis_pi4}(c) of the main text, where perfect cooperators receive zero payoff.

\subsection{Robustness: natural frequency heterogeneity}
\begin{figure}[ht!]
    \centering
    \includegraphics[clip, trim=3mm 3mm 5mm 7mm, width=0.7\columnwidth]{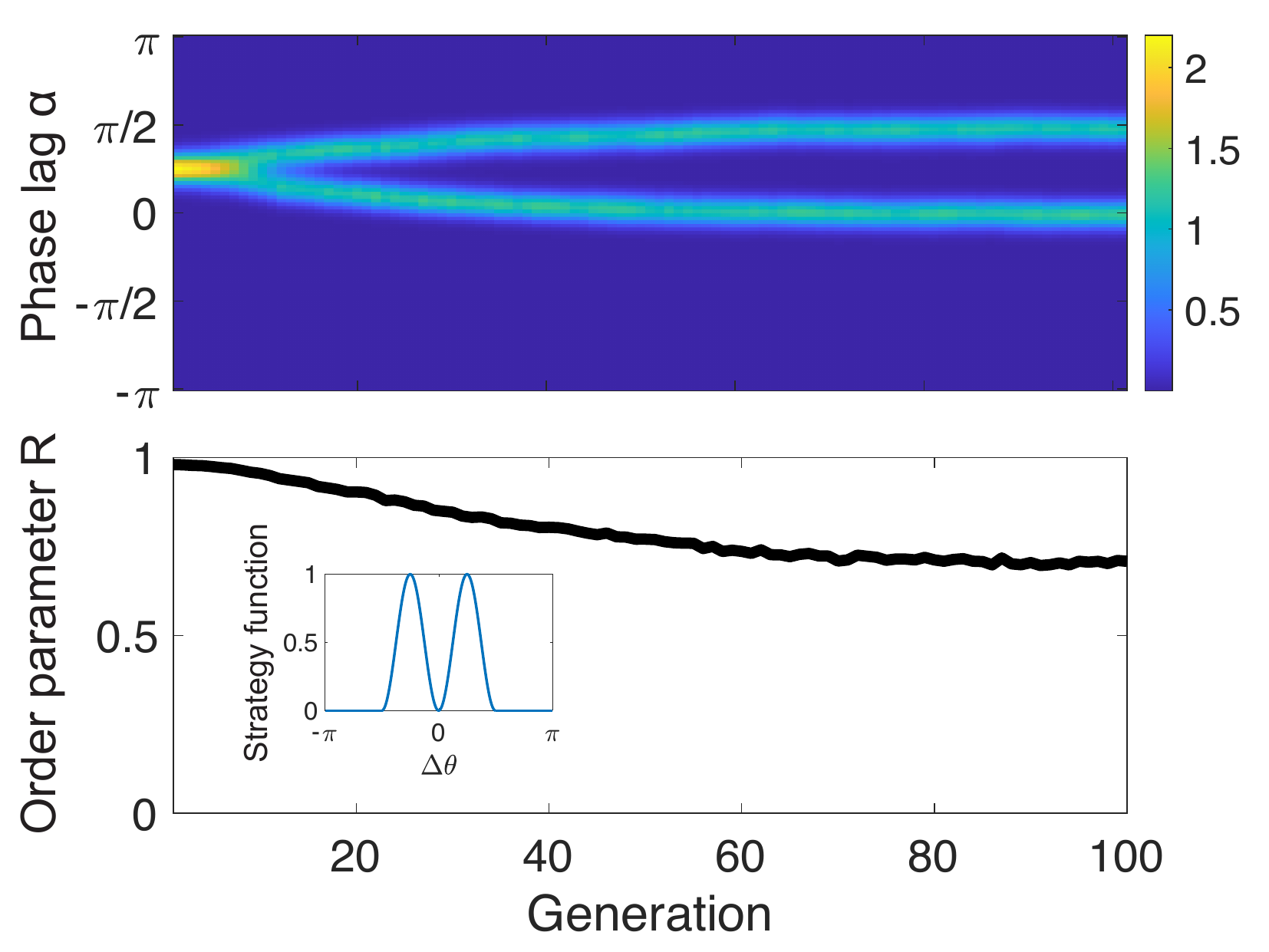}
    \caption{\textbf{Natural frequency heterogeneity.} To test the robustness of the simulation results to heterogeneity in the natural frequency distribution, instead of identical natural frequencies, we choose $\omega \sim \mathcal{N}(\mu_{\omega}, \sigma_{\omega})$ with $\mu_{\omega} = 0$ and $\sigma_{\omega} = 1$. The resulting population dynamics appears qualitatively unchanged when compared with Fig.~\ref{fig:population_dynamcis_pi4}(c), as long as the coupling strength is above a critical value. }
    \label{fig:population_dynamcis_heterogeneous} 
\end{figure}

We examine the robustness of the simulation results when the population has a heterogeneous natural frequency distribution $\omega \sim \mathcal{N}(\mu_{\omega}, \sigma_{\omega})$ with nonzero $\sigma_{\omega}$. As shown in Fig.~\ref{fig:population_dynamcis_heterogeneous} the results appear to exhibit the same qualitative behavior as those in Fig.~\ref{fig:population_dynamcis_pi4}(c) of the main text, where there is zero heterogeneity in natural frequency.

\end{document}